\documentclass[letterpaper, 10 pt, conference]{ieeeconf}%

\IEEEoverridecommandlockouts%
\usepackage{amsmath}%
\usepackage{amssymb}%

\usepackage[top=54pt, bottom=54pt, left=54pt, right=54pt]{geometry}
\usepackage{amsthm}
\usepackage{mathtools}
\usepackage{multirow}
\usepackage{rotating}
\usepackage{tabularx}
\usepackage{makecell}
\usepackage{array}
\usepackage[table]{xcolor}
\definecolor{hotpink}{RGB}{200, 50, 120}
\definecolor{green}{RGB}{0, 191, 40}
\usepackage{booktabs}

\usepackage{microtype}%
\usepackage{eso-pic}
\AddToShipoutPictureBG*{%
  \AtPageUpperLeft{\put(0,-30){\makebox[\paperwidth]{\parbox{0.86\paperwidth}{\centering\footnotesize
  \copyright~2026 IEEE. Personal use of this material is permitted. Permission from IEEE must be obtained for all other uses, in any current or future media, including reprinting/republishing this material for advertising or promotional purposes, creating new collective works, for resale or redistribution to servers or lists, or reuse of any copyrighted component of this work in other works.}}}}}

\newtheorem{theorem}{Theorem}
\newtheorem{lemma}{Lemma}

\newtheorem{remark}{Remark}

\title{\LARGE \bf
Robust Multi-Agent Pursuit with Preemptive Communication
via Reachability-Based Control Barrier Functions
}

\author{Madison Book and Justin Bradley%
\thanks{This work is supported in part by NSF-2514761, USDA 2023-67021-45990, and CNS-2221648.}%
\thanks{Authors are with the Department of Computer Science, NC State University, Raleigh, NC, USA
        {\tt\small \{mtbook, jmbradley\}@ncsu.edu}}%
}

\begin{document}

\maketitle
\thispagestyle{empty}
\pagestyle{empty}

\begin{abstract}
Safe multi-agent coordination typically assumes what real-world, deployed, resource-constrained robots cannot provide: continuous, perfect state sharing. Robust control strategies must be complemented with robust communication strategies to deliver resilient deployments. In this work, we propose a unified control/communication strategy that provides safe control to guarantee inter-agent safety under intermittent communication and demonstrate its effectiveness in a pursuit-evasion scenario. Each agent computes the reachable set of neighboring agents based on the last state information received, then solves for a safe control input based on control barrier functions to ensure agents avoid collision. We prove that inter-agent collision avoidance is guaranteed provided communication staleness remains bounded. As the reachable set inflates between state transmissions, each agent reasons about when to preemptively transmit its state information to neighboring agents based on novel metrics characterizing \emph{uncertainty risk} and \emph{kinematic risk}. We implement and simulate a variety of pursuit-evasion scenarios and show that preemptive communication reduces messages by 77.5\% compared to periodic and event-triggered baselines with zero collisions.
\end{abstract}

\section{INTRODUCTION}

Multi-agent safety guarantees based on Control Barrier Functions (CBFs) typically assume continuous state information \cite{borrmann_control_2015}. This assumption fails the moment bandwidth, power, or latency issues arise. On small, resource-constrained UAVs, this failure is the norm rather than the exception, leading to event-triggered safe control methods that rely on continuous state monitoring to trigger communication when deemed necessary \cite{grafe2022event}. In this work, we invert the usual communication trigger: rather than transmitting when an agent's own state crosses a threshold, agents transmit preemptively, when neighbors need the information to remain safe.

An alternative line of work models communication as discrete events \cite{breeden_2022_control} but treats control and communication as separate design problems leading to fragile real-world deployments \cite{muralidharan2021communication}. In reality, control must account for the uncertainty growth with time under stale information.

Together, these limitations suggest a need for safety-aware communication that incorporates onboard constraints while ensuring safety. Rather than traditional event-triggered mechanisms \cite{ge_2017_distributed}, which rely on self-state thresholds to transmit state, agents should transmit state when it is needed to preserve safety guarantees. This reframes communication as a safety mechanism.

We address this by combining reachability analysis with CBFs. Reachability analysis over-approximates the set of states an agent could occupy since its last communicated state \cite{bansal_hamilton-jacobi_2017}. CBFs enforce safety constraints by finding control inputs that ensure the state never leaves a safe set \cite{ames_2019_control}. Related to this work, they have been applied to pursuit-evasion \cite{shen_control_2024, davydov_pursuer_2021} and multi-agent navigation \cite{palani_connectivity_2025, chen_2020_guaranteed}. CBFs have also been combined with reachability analysis \cite{choi_robust_2021, kim2023safety}. But these efforts have left open the opportunity to co-design CBFs with communication in a safety-preserving manner.

\begin{figure}[h!]
    \centering
    \includegraphics[width=.9\columnwidth]{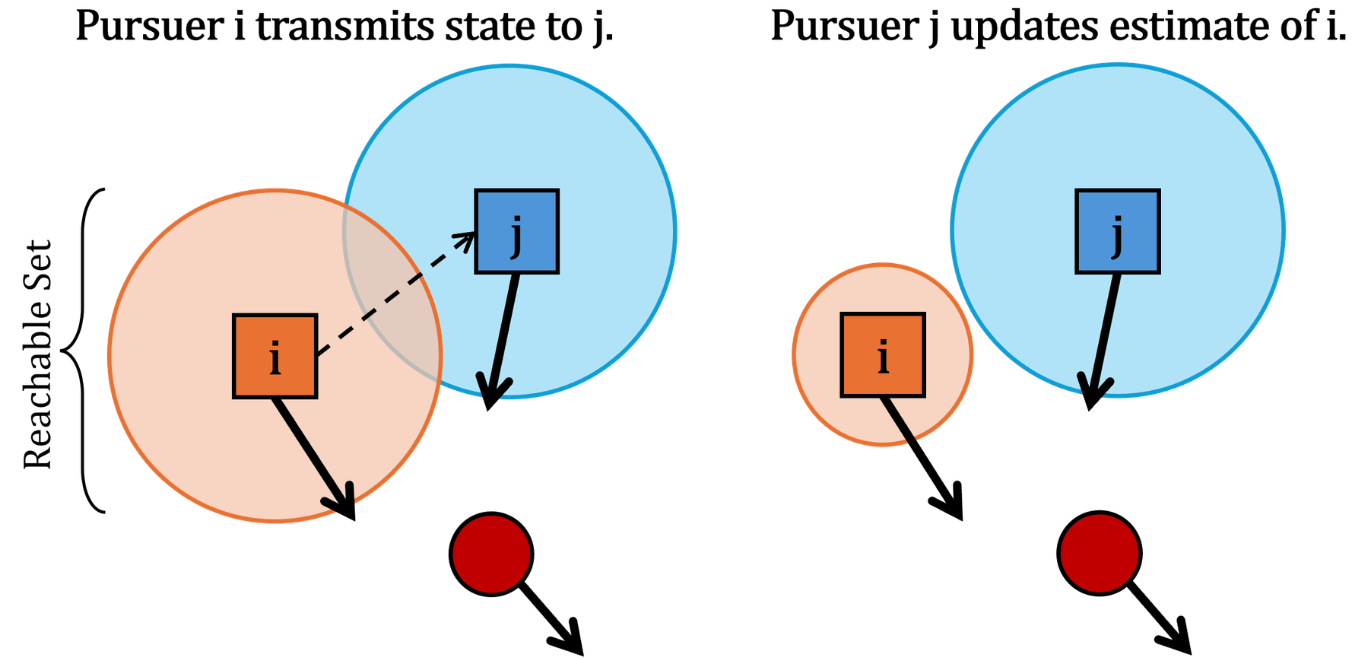}
    \caption{Demonstration of the preemptive communication mechanism. Pursuers $i$ and $j$ are represented as squares, their assigned evader is represented as a circle. On the left, pursuer $i$ identifies that it is at risk of collision with pursuer $j$ and therefore communicates its current state. Subsequently, $j$'s estimated reachable set for $i$ is reset, illustrated by the reachable set around $i$ shrinking on the right.}
    \label{fig:pe_header}
    \vspace{-10pt}
\end{figure}

As reachable sets expand with time, prolonged gaps in communication may lead to the quadratic programming (QP) controller becoming infeasible. To address this, we propose a preemptive communication mechanism where an agent transmits its state when it predicts that neighboring agents don't have enough information to remain safe. The result is agents that communicate with neighbors as needed, reducing overall communication bandwidth while still providing safety guarantees. This framework is illustrated in Figure \ref{fig:pe_header}.
 
We evaluate the proposed methods in a pursuit-evasion scenario with multiple cooperative pursuers and adversarial evaders. Pursuing agents must coordinate to capture the evaders while avoiding inter-agent collisions. The contributions of this paper are as follows:
\begin{enumerate}
    \item A novel preemptive communication mechanism based on predicted uncertainty and physical risk where agents transmit state when they predict neighboring agents need updated information.
    \item A reachability-aware control barrier function, providing formal safety guarantees under intermittent and stale information.
    \item A unified design strategy integrating preemptive communication and reachability-aware CBFs that accounts for onboard communication and computation limitations.
    \item Simulation results demonstrating that our unified strategy maintains safety and performance while reducing total message volume by 77.5\% against comparable strategies.
\end{enumerate}

\section{PROBLEM FORMULATION}

We consider a system of $n_p$ pursuers and $n_e$ evaders operating in $\mathbb{R}^3$. Pursuers and evaders are indexed by  $\mathrm{P} = \{1, \ldots, n_p\}$ and $\mathrm{E} = \{1, \ldots, n_e\}$ respectively. Each agent's state and dynamics are known and given by
\begin{equation}
    x= \begin{bmatrix} p \\ v \end{bmatrix}, \qquad
    \dot{x} = \begin{bmatrix} v \\ u \end{bmatrix},
    \label{eq:dynamics}
\end{equation}
where $p, v, u \in \mathbb{R}^3$ denote position, velocity and control input, respectively. We denote pursuer $i$'s state as $x_i$ and pursuer $i$'s assigned evader's state as $x_{e_i}$. We impose additional constraints satisfying pursuer maximum actuation bounds, $\|v_i\| \leq v_{\text{max}}$ and $\|u_i\| \leq u_{\text{max}}$, and assume that evaders have bounded acceleration.

For each neighboring pursuer $j$, agent $i$ maintains the last transmitted state $\xi_{ij} = \left(p_i(t_{ij}), v_i(t_{ij}), u_i(t_{ij})\right)$ and the last received state $\xi_{ji} = \left( p_j(t_{ji}), v_j(t_{ji}), u_j(t_{ji})\right)$, where $t_{ij}$ and $t_{ji}$ denote the respective transmission times. The staleness of each is given by $\tau_{ij} = t - t_{ij}$ and $\tau_{ji} = t - t_{ji}$.

\subsection{Assumptions}
\label{sec:assump}

Each pursuer $i \in \mathrm{P}$ has idealized knowledge of its own state $x_i$ and assigned evader's state $x_{e_i}$. We assume pursuer-evader assignments are fixed, known and computed using the Hungarian algorithm \cite{kuhn1955}. Each pursuer $i \in \mathrm{P}$ has stale, communicated knowledge of neighbors' states $x_j, j\neq i \in \mathrm{P}$. 

Pursuers communicate through an idealized directed, pairwise network. Rather than broadcasting at a fixed rate or based on self-state thresholds, each pursuer dynamically determines when to transmit its state to a neighbor based on physical risk and uncertainty. A heartbeat mechanism bounds the maximum time between transmissions and ensures that state information does not grow excessively stale. The full communication protocol is outlined in Section \ref{sec:comms}. 

\subsection{Objectives}

\subsubsection{Pursuer Objectives}

Each pursuer $i \in \mathrm{P}$ aims to maintain pairwise inter-pursuer safety for $j \neq i \in \mathrm{P}$ and capture its assigned evader $e_i \in \mathrm{E}$.

\subsubsection{Evader Objectives}

Evaders are adversarial. Each $e \in \mathrm{E}$ aims to move away from the nearest pursuer $i \in \mathrm{P}$.

\section{REACHABILITY-BASED CONTROL BARRIER FUNCTIONS}
\label{sec:cbfs}

\subsection{Reachability}
\label{sec:reach}

Since pursuer $i$ does not have continuous access to neighbor $j$'s state information, pursuer $i$ over-approximates the set of states that $j$ may occupy based on the last transmitted state $\xi_{ji}$. As $\tau_{ji}$ increases, the true position of $j$ may deviate according to unknown and bounded control inputs.

A pursuer $i$'s prediction of neighboring pursuer $j$'s current state is based on the last received state $\xi_{ji}$ and defined by
\begin{align}
    \hat{p}_j(t) &= p_j(t_{ji}) + v_j(t_{ji})\tau_{ji}, \\
    \label{eq:phat_j}
    \hat{v}_j(t) &= v_j(t_{ji}),
\end{align}
where $u_j(t_{ji})$ is omitted from predictions to reduce potential error and $j$'s reachable set size. Pursuer $i$ computes an over-approximated reachable sphere for each neighbor $j \neq i$, centered at $\hat{p}_j(t)$ and defined by
\begin{equation}
    \mathcal{S}_j(t) = \left\{\, q \in \mathbb{R}^3 \;\middle|\;
        \bigl\| q - \hat{p}_j(t) \bigr\| \leq r(\tau_{ji})
    \,\right\},
    \label{eq:sphere}
\end{equation}
where
\begin{equation}
    r(\tau_{ji}) = \frac{1}{2} u_{\max} \tau_{ji}^2
    \label{eq:radius}
\end{equation}
defines the radius of the sphere. The radius of the spherical reachable set grows quadratically with $\tau_{ji}$ and resets when new state information is received from $j$.

\begin{lemma}[Containment]
$p_j(t) \in \mathcal{S}_j(t) \ \forall t \geq t_{ji}$.
\end{lemma}
\begin{proof}
The true position $p_j$ is:
\[
  p_j(t) = p_j(t_{ji}) + v_j(t_{ji})\tau_{ji} + \int_0^{\tau_{ji}}\!\int_0^s u_j(\theta)\,d\theta\,ds.
\]
Thus, the position error $e_p(t) = p_j(t) - \hat{p}_j(t)$ satisfies
\begin{align*}
  \|e_p(t)\| &= \left\|\int_0^{\tau_{ji}}\!\int_0^s u_j(\theta)\,d\theta\,ds\right\|
               \le u_{\max}\int_0^{\tau_{ji}} s\,ds \\
             &= \tfrac{1}{2}u_{\max}\tau_{ji}^2 = r(\tau_{ji}).
\end{align*}
Since $\|p_j(t) - \hat{p}_j(t)\| \le r(\tau_{ji})$, it ensues that $p_j(t) \in \mathcal{S}_j(t)$ for all $t \ge t_{ji}$.
\end{proof}

\subsection{Control Barrier Functions}
\label{sec:qp}

\subsubsection{Inter-Pursuer Safety}

We impose a hard constraint for inter-pursuer pairwise safety. For each pursuer pair $(i, j) \in \mathrm{P}$, safety is defined by
\begin{equation}
    \|p_i - p_j\| \geq d_{\text{safe}}, \quad \forall\, i \neq j.
    \label{eq:collision}
\end{equation}
where $d_{\text{safe}} > 0$ is the required minimum separation between pursuers. Since $p_j$ is unknown, we use $j$'s reachable set $\mathcal{S}_j$ and predicted position $\hat{p}_j$ to define the barrier function
\begin{equation}
    h_{ij}(x_i, t) \coloneq \|p_i(t) - \hat{p}_j(t)\|^2 - \tilde{r}_j(\tau_{ji})^2, \quad \forall\, i \neq j
\end{equation}
where $\tilde{r}_j$ uses inflated radius $\tilde{r} = r + d_{\text{safe}}$. The safe set $\mathcal{C}_{ij} = \{x \mid h_{ij}(x) \geq 0\}$ is satisfied while pursuer $i$ is at least $d_{\text{safe}}$ outside of $j$'s spherical reachable set.

\subsubsection{Sensing Range Maintenance}

We impose a soft constraint for pursuer-evader sensing range maintenance. For each pursuer $i \in \mathrm{P}$ and assigned evader $e_i \in \mathrm{E}$ sensing range maintenance is defined as
\begin{equation}
    \|p_i - p_{e_i}\| \leq \kappa,
    \label{eq:sensing}
\end{equation}
where $\kappa > 0$ is the sensing radius. For each pursuer $i$ and assigned evader $e_i$, define the barrier function
\begin{equation}
    h_{ie_i}(x_i) \coloneq \kappa^2 - \|p_i(t) - p_{e_i}(t)\|^2,
    \label{eq:sensing_cbf}
\end{equation}
with safe set $\mathcal{C}_{ie_i} = \{x \mid h_{ie_i}(x) \geq 0\}$. The constraint is enforced softly via a slack variable $\delta \geq 0$ and penalty weight $w_s \gg 1$ in the QP controller. The soft constraint ensures priority of inter-pursuer safety over sensing range maintenance.

\subsubsection{High Order Control Barrier Functions}

Since $h_{ij}$ and $h_{ie_i}$ each have relative degree two with respect to control input, we employ the HOCBF conditions \cite{xiao_2022_high}.
\begin{align}
    \psi_{ij} &= \dot{h}_{ij}(x_i, t) + \alpha_0(h_{ij}(x_i,t)), \\
    \psi_{ie_i} &= \dot{h}_{ie_i}(x_i) + \alpha_2(h_{ie_i}(x_i))
\end{align}
where $\alpha_0(x) = a_0x, \alpha_2(x) = a_2x$ and $a_0, a_2> 0$ to enforce
\begin{align}
    \omega_{ij}(x_i, t) &= \dot{\psi}_{ij}(x_i,t) + \alpha_1(\psi_{ij}(x_i,t)) \geq 0, \label{eq:hocbf1}\\
    \omega_{ie_i}(x_i) &= \dot{\psi}_{ie_i}(x_i) + \alpha_3(\psi_{ie_i}(x_i)) \geq 0,
\end{align}
where $\alpha_1(x) = a_1x, \alpha_3(x) = a_3x$ and $a_1, a_3> 0$. Therefore, $\omega$ expands to
\begin{align}
    \omega_{ij}(x_i,t) &= \ddot{h}_{ij}(x_i,t) + k_1 \dot{h}_{ij}(x_i,t) + k_0h_{ij}(x_i,t), \\
    \omega_{ie_i}(x_i) &= \ddot{h}_{ie_i}(x_i) + k_3 \dot{h}_{ie_i}(x_i) + k_2h_{ie_i}(x_i),
    \label{eq:hocbf}
\end{align}
where $k_0 = a_0a_1, k_1 = a_0+a_1, k_2 = a_2a_3, k_3 = a_2+a_3$ such that $k_1^2 > 4k_0$ and $k_3^2 > 4k_2$ are the controller gains. 
The set of admissible control inputs satisfying hard safety constraints is therefore
\begin{equation}
    S = \bigl\{\, u_i \;\big|\; \omega_{ij}(x_i) \geq 0, \;
    \|u_i\| \leq u_{\max} \bigr\}
    \label{eq:safeset}
\end{equation}
satisfying \eqref{eq:hocbf1} guarantees forward invariance of the inter-pursuer safety set $\mathcal{C}_{ij}$ as demonstrated in Theorem \ref{thm}. Since $\omega_{ie_i}$ is a soft constraint, it is not required to satisfy safety.

\subsubsection{QP Controller}

During each control loop, pursuer $i$ solves the quadratic program
\begin{align}\label{eq:qp}
\begin{split}
    u_i^* = \arg\min_{u_i,\delta \ge 0} \quad 
    & \|u_i - u_{i,\mathrm{nom}}\|^2 + w_s \delta^2 \\
    \text{s.t.} \quad 
    & \omega_{ij}(x_i,t) \ge 0, \;\forall j \neq i, \\
    & \omega_{ie_i}(x_i) + \delta \ge 0, \\
    & \|u_i\| \le u_{\max}.
    \end{split}
\end{align}
where $u_i^*$ is the optimal control input and $u_{i,\mathrm{nom}}$ points towards the assigned evader's position $p_{e_i}$.

\begin{theorem}[Inter-Pursuer Collision Avoidance]
\label{thm}
If $h_{ij}(x_i, 0) \geq 0$, $\dot{\psi}_{ij}(x_i,t) + a_1\psi_{ij}(x_i,t) \geq 0$, $a_0,a_1>0$, $\forall t \ge 0$, and $\psi_{ij}(x_i,t_k) \geq 0$ at each communication event $t_k$, then $\|p_i(t) - p_j(t)\| \ge d_{\text{safe}}$ for all $t \ge 0$.
\end{theorem}
\begin{proof}
    \label{proof}
    \textit{Step 1: Show $\psi_{ij}(x_i,t) \ge 0 \quad \forall t \geq 0$.} The constraint $\tfrac{d\psi_{ij}}{dt} + a_1\psi_{ij} \ge 0$  gives $\tfrac{d\psi_{ij}}{\psi_{ij}} \ge -a_1\,dt$. Integrating yields \[
    \psi_{ij}(x_i,t) \ge \psi_{ij}(x_i,0)\,e^{-a_1 t}.
    \] If $\psi_{ij}(x_i,0) \ge 0$, then it follows that $\psi_{ij}(x_i,t) \ge 0$ $\forall t \ge 0$.

    \textit{Step 2: Show $h_{ij}(x_i, t) \geq 0 \quad \forall t \geq 0$.} Since $\psi_{ij}(x_i, t) = \dot{h}_{ij}(x_i,t) + a_0 h_{ij}(x_i,t) \ge 0 \quad \forall t \geq 0$, we obtain $\tfrac{dh_{ij}}{h_{ij}} \ge -a_0\,dt$. Integrating yields \[
    h_{ij}(x_i,t) \ge h_{ij}(x_i,0)\,e^{-a_0 t}.
    \] If $h_{ij}(x_i,0) \ge 0$, then it follows that $h_{ij}(x_i,t) \ge 0$ $\forall t \ge 0$.

    \textit{Step 3: Show $\|p_i - p_j\| \geq d_{\text{safe}}$ when $h_{ij}(x_i, t) \geq 0$.} $h_{ij} \geq 0$ enforces $\|p_i(t) - \hat{p}_j(t)\| \geq \tilde{r}(\tau_{ji})$. From Lemma 1, the true position of $j$ satisfies $\|p_j(t)-\hat{p}_j(t)\| \leq r$. Using the triangle inequality, we obtain
    \begin{align*}
        \|p_i(t) - p_j(t)\|
        &= \|(p_i - \hat{p}_j) + (\hat{p}_j - p_j)\| \\
        &\geq \|p_i - \hat{p}_j\| - \|p_j - \hat{p}_j\| \\
        &\geq \tilde{r}(\tau_{ji}) - r(\tau_{ji}) \\
        &= \bigl(\tfrac{1}{2}u_{\max}\tau_{ji}^2 + d_{\mathrm{safe}}\bigr) - \tfrac{1}{2}u_{\max}\tau_{ji}^2
        = d_{\mathrm{safe}}.
    \end{align*}
    As such, $\|p_i(t) - p_j(t)\| \geq d_{\text{safe}}$ $\forall t \geq 0$.

    The arguments in Steps 1-3 hold for any communication interval $[t_k, t_{k+1})$, as $\hat{p}_j$ and $r(\tau_{ji})$ evolve continuously through $\tau_{ji} = t - t_k$. We now show safety is preserved at and after each communication event at $t_k$.

    \textit{Step 4: Show $h_{ij}(x_i, t_k) \geq 0$ and $\psi_{ij}(x_i, t_k) \geq 0$ hold at each communication event $t_k$.} When pursuer $i$ receives $j$'s state information at time $t_k$, by hypothesis $\psi_{ij}(x_i, t_k) \geq 0$. Furthermore, at time $t_k$, $\tau_{ji} = 0$, therefore $r(0) = 0$ and $\tilde{r} = d_{\text{safe}}$. The barrier function now evaluates to
    \begin{equation}
        h_{ij}(x_i, t_k) = \|p_i(t_k)-\hat{p}_j(t_k)\|^2 - d_{\text{safe}}^2
    \end{equation}
    where $p_j(t_k) = \hat{p}_j(t_k)$, since prediction error is zero at the reset. Since $p_i$ and $p_j$ are continuous and Step 3 guarantees $\|p_i(t) - p_j(t)\| \geq d_{\text{safe}}$ on $[t_{k-1}, t_k)$, taking the limit $t \to t_k^-$ yields $h_{ij}(x_i, t_k) \geq 0$. Therefore, the conditions required by Steps 1 and 2 are satisfied at $t_k$. 

    \textit{Step 5: Show $\|p_i(t) - p_j(t)\| \geq d_{\text{safe}}$ $\forall t \geq 0$.} Suppose, as stated in Step 3 above, that $\|p_i(t) - p_j(t)\| \geq d_{\text{safe}}$ holds $\forall t\in[t_k,t_{k+1})$. By Step 4, $h_{ij}(x_i,t_{k+1})\geq 0$ and $\psi_{ij}(x_i,t_{k+1})\geq 0$ so the hypotheses of Steps 1-3 are satisfied at $t_{k+1}$. Steps 1-3 then guarantee that $\|p_i(t) - p_j(t)\| \geq d_{\text{safe}}$ $\forall t\in[t_{k+1},t_{k+2})$. By induction, $\|p_i(t) - p_j(t)\| \geq d_{\text{safe}}$ $\forall t\geq 0$.
\end{proof} 

\section{PREEMPTIVE COMMUNICATION}
\label{sec:comms}

\begin{remark}
Theorem 1 guarantees safety provided $\omega_{ij}(x_i,t_k)\geq 0$ can be enforced at each time step, keeping the QP \eqref{eq:qp} feasible. QP feasibility is threatened as $\tau_{ij}$ grows (i.e., the reachable set inflates, reducing control authority). The objective of the proposed communication strategy is to preemptively transmit state information to preserve QP feasibility while minimizing unnecessary communication.
\end{remark}

In the proposed decentralized pursuit-evasion scenario, pursuers must coordinate and maintain safety guarantees under intermittent and stale neighbor state information. Traditional communication strategies involve agents transmitting state information at fixed intervals or self-state changes. These methods do not account for cooperative agents that are in physically risky situations or do not have enough information about a teammate to operate safely. To address this, we propose a preemptive communication strategy, where pursuer $i \in \mathrm{P}$ transmits its state to $j \neq i \in \mathrm{P}$ when the estimated risk between them exceeds a threshold value. Agent $i$ transmits its state to a neighbor $j$ when $i$ predicts they are at kinematic risk of collision with $j$ or that $j$ doesn't have enough information about $i$ to operate safely.

Define the relative quantities $p_{ij} = p_i(t) - \hat{p}_j(t)$, $v_{ij} = v_i(t) - \hat{v}_j(t)$, $u_{ij} = u_i(t) - u_j(t_{ji})$, where $\hat{p}_j(t)$ and $\hat{v}_j(t)$ are computed using $\xi_{ji}$ and $\tau_{ji}$. Define $d_{ij} = \|p_{ij}\|$ and unit direction $n_{ij} = p_{ij}/d_{ij}$. To determine when to communicate, each pair of pursuers $(i, j)$ compute a preemptive risk score defined by
\begin{equation}
    \rho_{ij} = 1 - (1 - \rho_k)(1 - \rho_u),
    \label{eq:Rij}
\end{equation}
where $\rho_u$, \emph{uncertainty risk}, quantifies uncertainty due to staleness, and $\rho_k$, \emph{kinematic risk}, quantifies physical collision risk. This ensures that the estimated risk $\rho_{ij}$ is high when either uncertainty or kinematic risk is high, signaling agent $i$ to transmit state information. Pursuer $i$ transmits its current state to $j$ when
\begin{equation}
    \rho_{ij} \geq \mathcal{P} \quad \text{or} \quad \tau_{ij} \geq \mathcal{T},
    \label{eq:trigger}
\end{equation}
where $\mathcal{P} \in [0, 1)$ is the risk threshold and $\mathcal{T}$ is the maximum inter-communication interval. Smaller values of $\mathcal{P}$ result in more frequent, conservative communication, while larger values reduce communication at the expense of increased uncertainty and physical risk. $\mathcal{T}$ serves as a heartbeat mechanism to bound staleness regardless of risk and ensure that communication failures between agents can be detected. Inter-agent safety requires a finitely bounded $\mathcal{T}$. Both $\mathcal{P}$ and $\mathcal{T}$ can be tuned according to system constraints and performance.

The uncertainty risk $\rho_u$ measures pursuer $i$'s perception of pursuer $j$'s uncertainty about pursuer $i$. As staleness grows, this uncertainty inflates the reachable set used by $j$'s safety controller, eventually threatening QP feasibility. To quantify this, we define uncertainty risk as
\begin{equation}
    \rho_u = \frac{r_i(\tau_{ij})}{r_i(\tau_{ij}) + d_{ij}}.
    \label{eq:Ru}
\end{equation}
where $r_i(\tau_{ij})$ is the radius of $i$'s reachable set based on its last transmitted state to $j$. As such, $\rho_u \to 1$ when the reachable set uncertainty is large relative to predicted separation and $\rho_u \to 0$ when that uncertainty is small.

Conversely, the kinematic risk $\rho_k$, measures the relative motion of agents $i$ and $j$. Despite recent state information, pursuers may still be in situations where inter-agent safety is threatened by converging trajectories. To quantify this, we define kinematic risk as
\begin{equation}
    \rho_k =  1 -\prod_{n \,\in\, \{d,v,u\}} (1 - \beta_n\gamma_n), \quad \beta_n \in [0, 1),
    \label{eq:Rk}
\end{equation}
where $\gamma_d$, $\gamma_v$, and $\gamma_u$ represent proximity, velocity approach, and control input approach risks, respectively. This ensures the kinematic risk $\rho_k$ is high when any component $\beta_n\gamma_n$ is high. The weights satisfy $\beta_d > \beta_v > \beta_u$ to reflect the importance of current proximity over anticipated trajectory. The kinematic risk factors are defined by

\begin{equation}
    \gamma_d = 1 - \left(\frac{d_{ij} - d_{\mathrm{safe}}}{d_{ij}}\right)^{\!2},
    \label{eq:gd}
\end{equation}

\begin{equation}
    \gamma_v = \max\!\left(0,\ \frac{-n_{ij}^T v_{ij}}{2v_{\text{max}}}\right),
    \label{eq:gv}
\end{equation}

\begin{equation}
    \gamma_u = \max\!\left(0,\ \frac{-n_{ij}^T u_{ij}}{ 2u_{\text{max}}}\right).
    \label{eq:gu}
\end{equation}

The proximity risk satisfies $\gamma_d \to 1$ as $d_{ij} \to d_{\text{safe}}$ and $\gamma_d \to 0$ as the predicted distance increases. The velocity and control input approach risks, $\gamma_v$ and $\gamma_u$, increase as neighboring pursuers move or accelerate toward one another. The projected closing velocity and control input, $-n_{ij}^T v_{ij}$ and $-n_{ij}^T u_{ij}$, are positive when pursuers are converging and bound to $0$ otherwise. Both are normalized by $2v_{\text{max}}$ and $2u_{\text{max}}$ to ensure $\gamma_v, \gamma_u \in [0, 1]$. The components jointly capture current separation and predicted trajectories to measure pairwise collision risk.

The proposed preemptive strategy assumes homogeneous actuation constraints. For heterogeneous systems, $\gamma_v$ and $\gamma_u$ need to consider agent-specific actuation constraints.

\section{RESULTS}

In this section, we evaluate our proposed framework in a decentralized pursuit-evasion setting with multiple pursuers and evaders. We present the effectiveness of the preemptive communication strategy against baseline communication models. 
\begin{figure}[ht]
    \centering
    \includegraphics[width=.9\columnwidth]{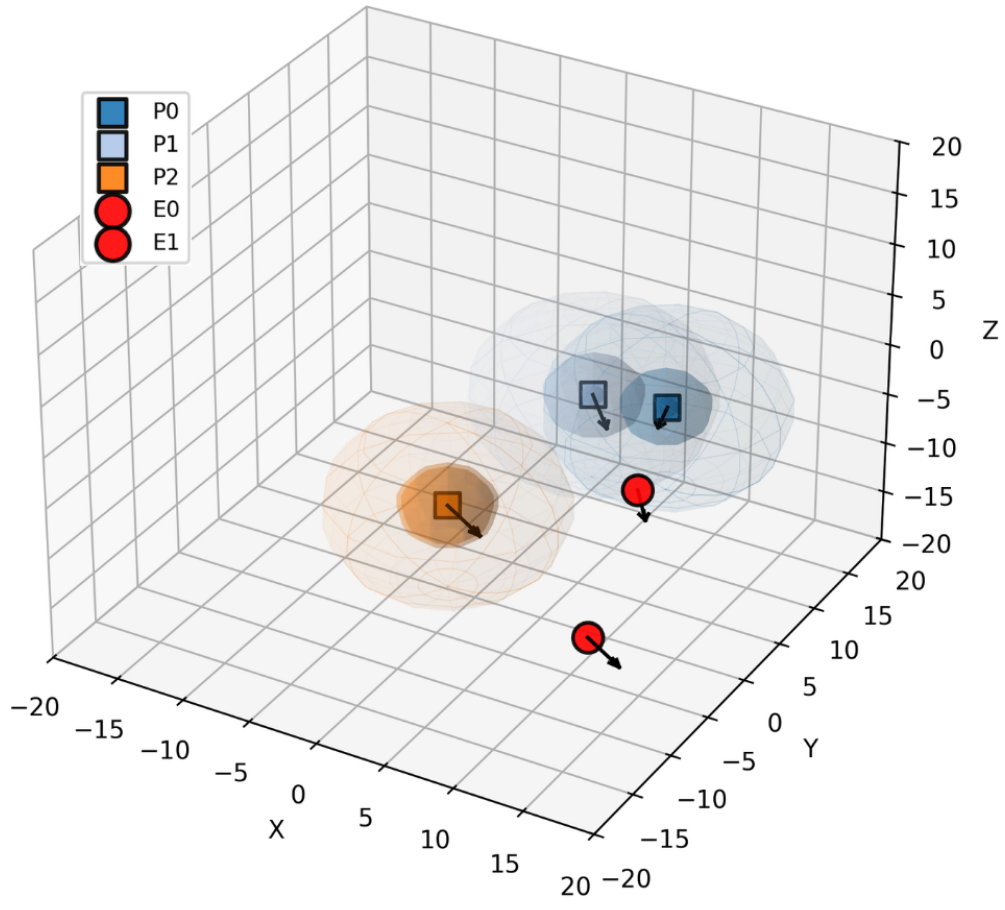}
    \caption{Snapshot of a simulation with 3 pursuers and 2 evaders. Pursuers are represented by squares, evaders by circles. Arrows indicate current velocity direction. Faint large spheres surrounding pursuers represent evader sensing range. Darker spheres represent predicted reachable sets.}
    \label{fig:pe_sim}
\end{figure}
\begin{figure}[ht]
    \centering
    \includegraphics[width=.9\columnwidth]{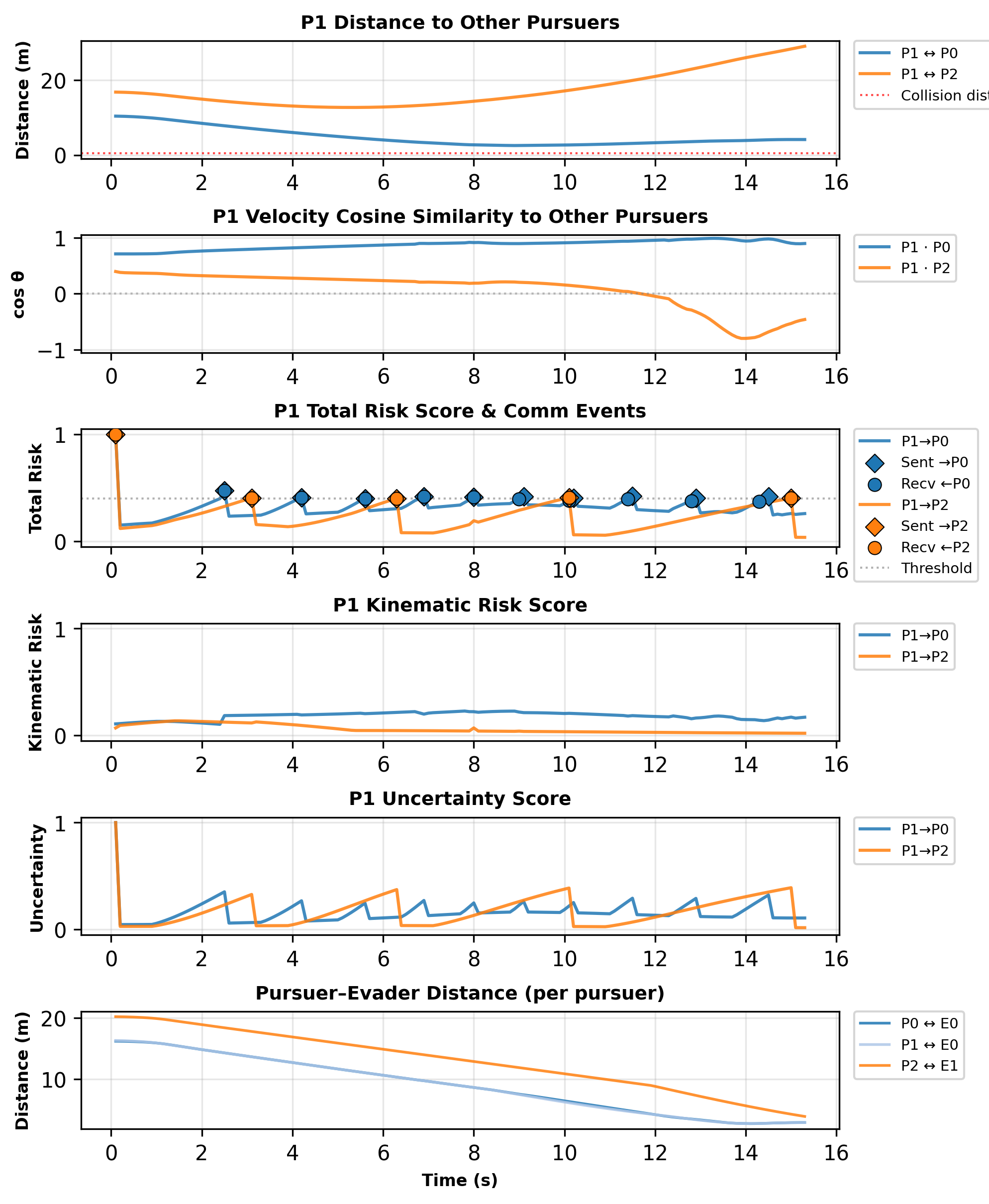}
    \caption{Simulation metrics for the 3 pursuer 2 evader case shown in Figure \ref{fig:pe_sim}. Distance, velocity, total risk, kinematic risk, and uncertainty risk plots follow pursuer 1 in Figure \ref{fig:pe_sim}. The preemptive communication mechanism ensures pursuers 1 and 0 communicate more frequently than pursuers 1 and 2 as they are at higher risk of collision. The pursuer-evader distance graph shows pursuers closing on and capturing their assigned evaders.}
    \label{fig:pe_graphs}
    \vspace{-20pt}
\end{figure}

The simulation consists of multiple cooperative pursuers attempting to capture multiple adversarial evaders. Pursuers follow the dynamics outlined in (\ref{eq:dynamics}) and use the safe controller outlined in Section \ref{sec:qp} to enforce inter-agent collision avoidance. Agents are randomly initialized in the environment subject to the specified safe distance, and pursuer-evader assignments are computed using the Hungarian algorithm described in Section \ref{sec:assump}. Pursuers have idealized access to the state of their evader and intermittent access to the states of neighboring pursuers. As such, pursuers maintain predicted state estimates for neighbors that are propagated according to the reachable set described in Section \ref{sec:reach}.

Figure \ref{fig:pe_sim} illustrates a representative snapshot of the simulation in the 3 pursuer 2 evader case. 

Figure \ref{fig:pe_graphs} outlines the metrics for the 3 pursuer 2 evader case illustrated in Figure \ref{fig:pe_sim}. The distance, velocity, risk score, kinematic risk, and uncertainty risk plots follow pursuer 1 as identified in Figure \ref{fig:pe_sim}. These plots depict the preemptive communication mechanism described in Section \ref{sec:comms}. Pursuer 1 communicates with pursuer 0 more frequently than pursuer 2 as they are closer and therefore at higher risk of collision. The pursuer-evader distance graph conveys the pursuers closing on and capturing their assigned evaders.

We compare the proposed preemptive communication strategy against baseline communication mechanisms. These are outlined as follows:
\begin{enumerate}
    \item None. No communication occurs between agents. Pursuers chase their assigned evader without the controller enforcing safety.
    \item Full. Pursuers share state information at each time step and use the described safe controller.
    \item Periodic. Pursuers share state information at frequencies 1 and 5 Hz and use the described safe controller. 
    \item Event Triggered. Pursuers share state information when the change in their state exceeds a predefined threshold relative to the prior transmission and use the described safe controller.
\end{enumerate}
\begin{table}[ht]
    \centering
    \caption{Simulation Parameters}
    \label{tab:sim_params}
    \begin{tabular}{@{}lll@{}}
    \toprule
    Description & Parameter & Value \\
    \midrule
    Time step & $\Delta t$ & 0.1 s \\
    Pursuer maximum velocity & $v_{\text{max}}$ & 2.0 m/s \\
    Pursuer maximum control & $u_{\text{max}}$ & 1.5 m/s$^2$ \\
    Safe distance & $d_{\text{safe}}$ & 0.5 m \\
    Pursuer's sensing range & $\kappa$ & 4 m \\
    HOCBF controller gains & $k_0, k_2$ & 2 \\
    HOCBF controller gains & $k_1, k_3$ & 4 \\
    QP penalty weight & $w_s$ & 1000 \\
    Preemptive distance weight & $\beta_d$ & 0.6 \\
    Preemptive velocity weight & $\beta_v$ & 0.25 \\
    Preemptive control weight & $\beta_u$ & 0.15 \\
    Preemptive risk threshold & $\mathcal{P}$ & 0.4 \\
    Preemptive maximum interval & $\mathcal{T}$ & 5 s \\
    \bottomrule
    \end{tabular}
    \vspace{-15pt}
\end{table}

The simulation parameters are outlined in Table \ref{tab:sim_params}. The $\mathcal{T}$ threshold was decided using real deployment data, where 5 seconds was determined to be the maximum amount of time to risk identifying a lost agent.

\begin{table*}[ht]
\centering
\caption{Communication Mechanism Performance Comparison}
\label{tab:comm_metrics}
\begin{tabular}{@{}llllllll@{}}
\toprule
\textbf{Scenario} & \textbf{Metric} & \textbf{None} & \textbf{Periodic 1Hz} & \textbf{Periodic 5Hz} & \textbf{Event Triggered} & \textbf{Full} & \textbf{Preemptive} \\
\midrule

\multirow{4}{*}{\textbf{8P, 8E}}
  & Avg Total Msgs     & 0.00 & \cellcolor{red!20}817.60   & \cellcolor{red!40}3183.60  & \cellcolor{red!30}1356.18  & \cellcolor{red!60}8536.08  & \cellcolor{green!20}259.82 \\
  & Msgs/Agent/s       & 0.00 & \cellcolor{red!20}6.71     & \cellcolor{red!40}26.14    & \cellcolor{red!30}11.13    & \cellcolor{red!60}70.01    & \cellcolor{green!20}2.12   \\
  & Total Collisions   & \cellcolor{red!50}10 & 0.0 & 0.0 & 0.0 & 0.0 & 0.0 \\
  & Avg Total Time     & 15.34s & 15.24s & 15.22s & 15.23s & 15.24s & 15.33s \\
\midrule

\multirow{4}{*}{\textbf{12P, 6E}}
  & Avg Total Msgs     & 0.00 & \cellcolor{red!20}1632.84  & \cellcolor{red!40}6306.96  & \cellcolor{red!30}2829.75  & \cellcolor{red!60}16675.56 & \cellcolor{green!20}604.80 \\
  & Msgs/Agent/s       & 0.00 & \cellcolor{red!20}10.64    & \cellcolor{red!40}41.45    & \cellcolor{red!30}18.67    & \cellcolor{red!60}110.02   & \cellcolor{green!20}3.95   \\
  & Total Collisions   & \cellcolor{red!50}80 & 0.0 & 0.0 & 0.0 & 0.0 & 0.0 \\
  & Avg Total Time     & 12.70s & 12.79s & 12.68s & 12.63s & 12.63s & 12.77s \\
\midrule

\multirow{4}{*}{\textbf{16P, 4E}}
  & Avg Total Msgs     & 0.0  & \cellcolor{red!20}2596.80  & \cellcolor{red!40}10202.40 & \cellcolor{red!30}4795.95  & \cellcolor{red!60}26455.20 & \cellcolor{green!20}1011.08 \\
  & Msgs/Agent/s       & 0.0  & \cellcolor{red!20}14.56    & \cellcolor{red!40}57.45    & \cellcolor{red!30}25.00    & \cellcolor{red!60}150.04   & \cellcolor{green!20}5.72    \\
  & Total Collisions   & \cellcolor{red!50}100 & 0.0 & 0.0 & 0.0 & 0.0 & 0.0 \\
  & Avg Total Time     & 10.87s & 11.12s & 11.10s & 11.99s & 11.02s & 11.04s \\
\bottomrule
\end{tabular}
\vspace{-15pt}
\end{table*}

\subsection{Discussion of Results}

Table \ref{tab:comm_metrics} presents a comprehensive performance comparison across three pursuit-evasion scenarios, including 8 pursuers and 8 evaders (8P8E), 12 pursuers and 6 evaders (12P6E), and 16 pursuers and 4 evaders (16P4E), over 100 trials each. Simulations focused on situations where pursuers were actively closing on evaders. Performance is evaluated along four axes: total messages transmitted, messages per agent per second, inter-pursuer collisions, and total capture time. The results consistently demonstrate that the proposed preemptive communication strategy achieves collision-free operation and competitive task performance at a fraction of the communication cost of all baseline methods. 

\subsubsection{Communication Efficiency}

The proposed preemptive mechanism achieves the lowest communication overhead of any strategy that enforces safety, across every scenario tested. Compared to the full communication strategy, the preemptive strategy sends 29.03$\times$ less messages/agent/second and reduces overall message volume by 96.5\% while maintaining comparable safety and capture performance.

The preemptive mechanism also outperforms the periodic baseline mechanisms across all scenarios. The preemptive strategy reduces overall message volume by 64.1\% compared to the 1 Hz strategy and 90.77\% compared to the 5 Hz strategy. The preemptive strategy sends 3.26$\times$ and 10.96$\times$ fewer messages/agent/s against the 1 Hz and 5 Hz strategies, respectively. Compared to the event-triggered mechanism, the preemptive strategy sends 4.78$\times$ fewer messages/agent/s, which reduces overall message volume by 79.43\%.

Across all scenarios and baseline communication strategies, the preemptive mechanism reduces total message volume by 82.2\%. Compared to the periodic and event-triggered mechanisms, the preemptive strategy reduces total message volume by 77.5\%. This reduction is not uniform over time. The preemptive communication strategy concentrates communication when agents are at risk of collision or predict that neighbors need information to operate safely. This effectively reframes communication as a mechanism for safety, where agents transmit their state when they predict their neighbors need information to operate safely.

\subsubsection{Collision Avoidance}

Inter-agent safety is the primary constraint in multi-agent systems, and it is strictly maintained by the communicating strategies across all scenarios. This demonstrates that the reachability-based HOCBF-QP controller provides a robust safety guarantee even under substantially reduced information availability.

When agents do not communicate, collisions are unavoidable. Collisions increase with agent count due to the higher density of close encounters as the swarm size scales.

\subsubsection{Total Capture Time}

In all cases, total capture time between the preemptive strategy and the communicating baselines are within a fraction of a second. This indicates that the 82.2\% reduction in message volume incurs no significant penalty in the total capture or task completion time.

\subsubsection{Scalability}

The advantages of the preemptive strategy compound with scale. Full and periodic strategies scale with agents, as their message rates are governed by the number of agent pairs and communication interval. The event-triggered strategy scales similarly, as more agents operating in a space leads to a higher frequency of significant self-state changes. 

In contrast, the preemptive mechanism's communication rate is governed by the situational risk, resulting in more favorable scaling as the number of agents in a system grows. 

\subsubsection{Summary}

The results demonstrate that our preemptive communication mechanism achieves the same performance as other baseline communicating methods, including safety and capture time, while significantly reducing the communication required. Across all scenarios and baseline communication strategies, the preemptive strategy reduces total message volume by 82.2\%, further indicating that safety-aware communication can significantly reduce overall communication while maintaining provable safety guarantees. 

\section{CONCLUSION}

This paper presented a preemptive communication strategy, where agents transmit state when they predict neighbors need information to operate safely. When combined with a reachability-based HOCBF-QP controller to enforce inter-agent safety under intermittent state information, the preemptive communication strategy maintained zero inter-agent collisions while reducing the total message volume by 82.2\% compared to baseline communication strategies. 

Future work will focus on more realistic reachable set formulations and deployment on SWaP-constrained UAV platforms to validate real-world effectiveness.

\bibliographystyle{IEEEtran}
\bibliography{references}

\end{document}